# YouSkyde: Information Hiding for Skype Video Traffic

Wojciech Mazurczyk, Maciej Karaś, Krzysztof Szczypiorski, Artur Janicki
Warsaw University of Technology, Institute of Telecommunications
Warsaw, Poland, 00-665, Nowowiejska 15/19

**Abstract.** In this paper a new information hiding method for Skype videoconference calls – YouSkyde – is introduced. A Skype traffic analysis revealed that introducing intentional losses into the Skype video traffic stream to provide the means for clandestine communication is the most favourable solution. A YouSkyde proof-of-concept implementation was carried out and its experimental evaluation is presented. The results obtained prove that the proposed method is feasible and offer a steganographic bandwidth as high as 0.93 kbps, while introducing negligible distortions into transmission quality and providing high undetectability.

**Key words:** information hiding, Skype, network steganography

## 1.   Introduction

The last few years have been marked by the proliferation of various multimedia services for Internet protocol (IP) networks: IP telephony, videoconferencing, Video on Demand (VoD), online games, IP-TV and so on, and this has significantly changed the way we communicate, work and entertain. From the network traffic perspective, these services typically exchange data in the form of streams, and the resulting traffic is characterized by quite a high volume of up to hundreds of packets per second. Also, not uncommonly the generated data is transmitted in an encrypted form. One of the most popular services of this kind is Skype, which not only allows IP telephony or videoconference calls to be conducted, but also chat messages to be exchanged and/or the sharing of files. Skype is a proprietary peer-to-peer (P2P) telephony service originally introduced in 2003 by the creators of the famous P2P file-sharing system, Kazaa (Niklas Zennström and Janus Friis). Currently, it is owned by Microsoft and in 2013 it was used by more than 280 million people every month [12]. In June 2014, the highest number of simultaneous users online was reported (about 80 million [11]), and recently Skype's share of the world's international telephone market has exceeded 40% [10].

 A variety of information hiding methods has been used throughout the ages by people. One of these methods is steganography, the main aim of which is to embed a secret message into a carefully selected carrier [17]. The youngest branch of this information hiding subfield is network steganography. To perform hidden communication, it utilizes network protocols and/or their relationships as the hidden data carrier. Network steganography is typically utilized to create a covert network channel; thus, it is intended to hide the very existence of the communication. Thus, any third-party observer would remain unaware of the presence of the steganographic exchange. This particular feature of network steganography makes it suitable for applications where stealth is of crucial importance [18]. We believe that the most important use of network steganography is to provide a means of clandestine communication to improve the user's privacy. Hence, information hiding techniques can be used as part of the tool, for example, to fight censorship in oppressive regimes. Skype is unquestionably well-fitted for covert communication purposes, mainly because of its popularity and traffic volume, and because, as mentioned earlier, any steganographic method requires a well-selected hidden data carrier to succeed. An ideal candidate for a carrier should: *(i)* be popular, in other words, usage of such a carrier should not be considered an anomaly in itself – the more such carriers are present and utilized on the network the better, because they mask the existence of hidden communication; *(ii)* the modification of the carrier related to the embedding of the steganogram should not be "visible" to the third party, who is unaware of the steganographic procedure. Thus, targeting Skype was a natural choice for steganographers as it fulfils both of these requirements.

 In general, every network steganographic method can be described by the following set of characteristics [2]:
 • *Steganographic bandwidth*, which refers to the amount of secret data that can be sent per unit of time when using a particular information hiding technique;
 • *Undetectability*, which is defined as the inability to detect secret data within a certain carrier;

- *Robustness*, which is defined as the amount of alteration that secret data can withstand without being destroyed.
- *Steganographic cost*, which indicates the distortion of the hidden data carrier caused by applying the steganographic method.

It must be emphasized that a good steganographic method should be as robust and as difficult to detect as possible, whilst offering the highest bandwidth. However, it should be noted that a fundamental trade-off is always necessary among these measures.

One can always question why apply information hiding techniques to Skype, which uses encryption to provide confidentiality for every type of message that is exchanged (text messages, voice signal, files). Our belief is that it is important for the following reasons:
- Firstly, discussion has arisen about whether Skype, whose calls were commonly believed to be very hard to wiretap, is providing lawful interception services to law enforcement agencies [4]. Snowden revealed that nine months after Microsoft bought Skype, the National Security Agency (NSA) had tripled the amount of Skype video calls being collected through the PRISM programme [9].
- Secondly, hidden communications need not necessarily be conducted in an end-to-end manner; that is, covert communication can be conducted using third-party Skype traffic.
- Finally, Skype is a proprietary and closed software; thus, it ultimately cannot be trusted.

This is why, in this paper we discuss how Skype video traffic can be utilized to provide a means of clandestine communication. Experiments performed on real Skype traffic have proved that the method we propose, YouSkyde, which relies on introducing intentional losses of video packets, is feasible and offers a high steganographic bandwidth and a reasonable steganographic cost under the terms of undetectability.

The rest of the paper is structured as follows: Section 2 describes related work on network steganography and on Skype steganography in particular; Section 3 presents a Skype video traffic analysis; Section 4 describes the functioning of YouSkyde in detail; Section 4 discusses the proof of concept implementation; Section 5 presents its experimental evaluation; finally, Section 6 concludes our work.

## 2. Related Work

As mentioned above, the typical user content exchanged using Skype is voice, video, text or files and the transmission is governed and data encapsulated within Skype proprietary protocols. For each of these types of data the resulting traffic can be subject to information hiding.

In general, steganographic methods that can be applied to network traffic can be broadly classified into two groups: storage and timing methods.

Firstly, the network covert channel can be created using steganographic techniques by modifying specific fields in the various protocol data units of Skype traffic. This includes network or transport protocols from TCP/IP stack (such as IP, TCP/UDP) or higher, application level protocols. However, as Skype is a proprietary service and the major part of the traffic is encrypted, the utilization of such methods is quite limited due to their potentially negative impact on the quality of the transmission.

Secondly, the covert channel can be realized through the use of information hiding methods that modify the time dependencies between packets in the stream. This is a more promising approach for Skype traffic because it potentially impacts less on the overall quality of the transmission, which is why in this study the main focus will be on these solutions. Methods for modifying packets in the stream are typically based on the following approaches:
- Influencing the sequence of the packets by assigning an agreed-upon order of packets during a predetermined period of time [15]. For example, sending packets in an ascending order could indicate a binary one, and a descending order a binary zero.
- Modifying inter-packet delays [13]; for example, where predetermined delays between two subsequent packets are used to send a single secret data bit. A similar concept is applied when different sending rates are utilized for the packets in the stream [14]. In a simplified scenario, one (the original) rate denotes a binary one, a second rate (e.g. achieved by delaying packets) means a binary zero.

- Introducing intentional losses [16]. This can be realized by, for example, skipping one sequence number while generating packets. From the receiver's perspective the detection of this "phantom" loss during a predetermined period of time means sending one bit of secret data.

The first Skype-dedicated method was introduced by Wang et al. [3]. The authors proposed embedding a 24-bit watermark into the encrypted stream (e.g. a Skype call) to track its propagation through the network, thus de-anonymizing it. The watermark was inserted by modifying the inter-packet delay for selected packets in the voice stream. The authors demonstrated that, depending on the watermark parameters chosen, they were able to achieve a 99% true positive and a 0% false positive rate, while maintaining good robustness and undetectability. However, they achieved a steganographic bandwidth of only about 0.3 bit/s, which is enough for the described application, but rather low for performing clandestine communication.

However, the first information hiding method that enables covert communication to take place using Skype traffic – SkyDe (Skype Hide) – has been recently proposed by Mazurczyk et al. [1]. SkyDe takes advantage of the high correlation between speech activity and packet size in Skype, and of the fact that no silence suppression mechanism is utilized. It is therefore possible to identify packets by silence (even if they are encrypted) and reuse their payloads to carry encrypted secret data. From the Skype service perspective, SkyDe relies on introducing intentional losses to the stream; however, their impact on the transmission quality is limited. Experimental results obtained using proof-of-concept implementation have proved that such an approach offers a high steganographic bandwidth of up to 1.8 kbps, while introducing almost no distortions to the Skype call (packets with silence do not influence the quality of the call and are highly undetectable).

The YouSkyde method proposed in this paper operates by reusing encrypted video packets in a videoconference call, by substituting payloads of carefully selected packets with encrypted secret data. From the Skype perspective it looks as if intentional losses are introduced. When compared with existing Skype-based information hiding solutions, it is characterized by a significantly higher steganographic bandwidth. Moreover, YouSkyde can be treated as a complementary method for SkyDe as the former can be applied to Skype video and the latter to voice traffic. However, it must be noted that if applied jointly the introduced losses could be additive.

## 3. Skype basics and video traffic analysis
### 3.1 Skype fundamentals

A Skype network forms a hierarchical P2P topology with a single centralized element (login server) that is created using two types of nodes [7]: *(i)* Ordinary Nodes (ONs) that can start and receive a call, send instantaneous messages and transfer files; *(ii)* special-purpose nodes, known as Super Nodes (SNs), that are responsible for helping ONs find and connect to each other within the Skype network. The login server is responsible for the authentication of ONs and SNs before they access the Skype network. Historically, every user's device with sufficient resources (in terms of CPU, bandwidth) and a public IP address could become an SN. However, it is believed that from 2012 Microsoft itself has hosted about 100,000 SNs [19], so the rest of the user nodes are ONs.

Typically, Skype offers two communication modes, namely: End-to-End (E2E) and End-to-Out (E2O). The former takes place between two Skype clients within the IP network, while the latter occurs if one of the endpoints is a Skype client within the IP network and the other is a PSTN phone (SkypeIn/SkypeOut services). In this paper, we focus solely on the Skype E2E mode.

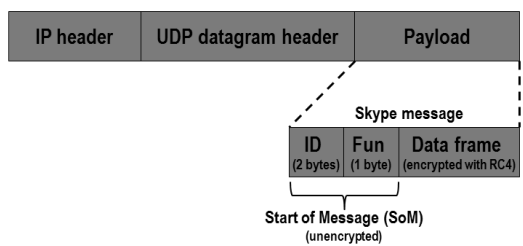

**Figure 1. UDP-based Skype message format.**

As already mentioned, Skype is based on proprietary protocols and makes extensive use of cryptography, so that neither the signalling messages nor the packets that carry voice data can be uncovered. Moreover, it utilizes various obfuscation and anti-reverse-engineering procedures. Thus, all the information about its traffic characteristics, protocols and behaviours comes from numerous measurement studies (e.g. [5], [6]). However it must be noted that they mostly focus on characterizing voice-call traffic.

Typically, the preferred, first-choice transport protocol for Skype is the User Datagram Protocol (UDP), which the traffic analysis in [6] showed is being used for about 70% of all calls. However, if Skype is unable to connect using UDP it falls back to TCP (this is true especially for audio streams). In this paper, we focus solely on UDP-based Skype video calls.

For TCP-based transport, the entire Skype message is encrypted, while in the case of UDP-based transport the beginning of each datagram's payload possesses an unencrypted header called the Start of Message (SoM). It is necessary to restore the sequence of packets that was originally transmitted to detect a loss, and to quickly distinguish the type of data that is carried inside the message. SoM consists of the following two fields [5] (Fig. 1): *(i)* ID *(2 bytes)* that is used to uniquely identify the message, which is randomly selected by the sender query and copied in the receiver reply; *(ii) Fun (1 byte)*, which describes the payload type. For example, values: 0x02, 0x03, 0x07 and 0x0f are typically used to indicate signalling messages (used during the login phase or for connection management). 0x0d indicates a DATA message that can contain encoded voice or video blocks, chat messages or fragments of files.

From the video traffic perspective, which is the core interest of the proposed steganographic method, after 2005 Skype utilized the True Motion VP7 codec [20] and from early 2011 version 5.5 moved to VP8. Both video codes were produced by On2 Technologies for Google. Thus, when Microsoft acquired Skype the video codec has been changed to H.264 [21].

It must be also noted that the Skype video call quality can be of the following standards (https://support.skype.com/en/faq/FA597/what-do-i-need-to-make-a-video-call):
- *Standard*, which means that the utilized video resolution will be 320 × 240 pixels and the video signal will be sent at a rate of 15 fps.
- *High-quality*, in which the resolution is 640 × 480 pixels and the frame rate is 30.
- *HD*, which is characterized by the highest resolution of 1280 × 720 pixels and the same rate of 30 fps.

As for the typical network steganography method, the general rule is the more traffic there is, the better the potentially resultant steganographic bandwidth for Skype video traffic. Thus it can be expected to be higher than for existing information hiding methods for audio streams.

**3.2 Skype video measurements test-bed and traffic analysis**

An analysis of Skype video traffic was performed to prove that the proposed steganographic method is feasible. For this, an experimental test-bed was set up in order to analyse the traffic and then to evaluate YouSkyde (Fig. 2). The test-bed included two Skype clients and a Linux-based application designed and developed by authors that could intercept Skype packets before they reached (for the transmitting side) and after they entered (for the receiving side) the network interface. Two instances of this application were synchronized using the Network Time Protocol (NTP) and were responsible for the generation of reports regarding the video packets' statistics.

The measurement procedure was as follows. First, on the transmitting side the virtual video device was created using *v4l2loopback* (https://github.com/umlaeute/v4l2loopback). This was then used as a video device for Skype and allowed to play the chosen video files as inputs. Second, *Gstreamer* (http://gstreamer.freedesktop.org), which is a virtual multimedia framework, was utilized as a server to transmit the video stream to a virtual video device created in the previous step. On the receiving side *SimpleScreenRecorder* (http://www.maartenbaert.be/simplescreenrecorder) was applied, which is an application that allows part of the computer screen to be captured. As we were unable to capture the received video signal directly from Skype, it was necessary to acquire the part of the screen where the video call was displayed. This of course meant the potential degradation of the calculated video quality. The captured traffic as well as the sent and received video sequences were then subjected to analysis. Video quality was assessed using MSU Video Quality Measurement Tool (VQMT) (http://compression.ru/video/quality_measure/info_en.html#start), which was designed for objective video signal quality evaluation. This tool was developed at the Graphics and Media

Lab at M.V. Lomonosov Moscow State University, Russia. It allows a wide spectrum of video quality metrics to be determined that belong to one of two groups: mathematically defined metrics or metrics that have similar characteristics to the Human Visual System (HVS). For the purposes of this paper, we decided to calculate the Peak Signal-to-Noise Ratio (PSNR) that belongs to the first group and is currently the most popular and widely used metric, and the Multi Scale-Similarity Index Metric (MS-SSIM) and Video Quality Metric (VQM) as the representatives of the second group. VQM, especially, has been proven to be well correlated with subjective video quality assessment and has been adopted by ANSI as an objective video quality standard.

For the video testing sequence we decided to utilize video from the Video Trace Library of Arizona State University, USA (http://trace.eas.asu.edu/yuv/akiyo/). The sequence is of a news presenter speaking, which reflects the format of a typical Skype video call: a static background and an (almost static) upper part of the person in the middle of the screen. The video file was resized, replicated and merged to represent a few minutes of a videoconference call.

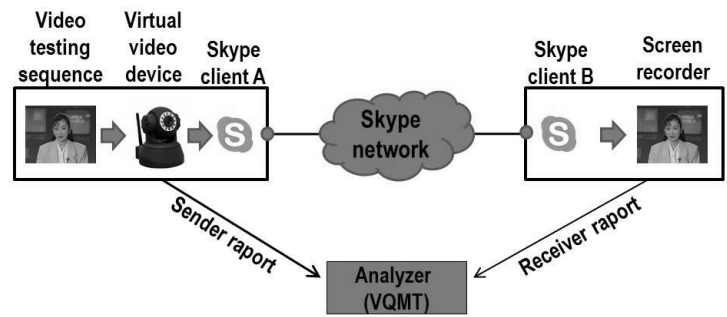

**Figure 2. Experimental Skype and YouSkyde test-bed.**

Using the test-bed presented in Fig. 2 and the abovementioned tools and quality metrics, a number of measurements were carried out on the Skype traffic. The averaged results for this study are presented below. It should also be noted that we limited our experiments to standard Skype mode. Therefore, the input video stream is transmitted to the virtual video device at a rate of 15 fps rate and $320 \times 240$ pixels resolution; the duration of the single video call was between 3 and 5 minutes.

For further study, it is vital to establish the reference quality of the received video signal in the test-bed presented in Fig. 2 without also applying the steganographic method. This will later allow the impact of the information hiding technique on the transmission quality from user's perspective to be assessed. Table 1 presents the experimental results obtained for the chosen video quality metrics.

**Table 1.** Quality metrics for reference video measurements

| Quality metric | Average | Std. dev. |
| --- | --- | --- |
| PSNR [dB] | 29.82 | 0.86 |
| VQM | 2.38 | 0.19 |
| MS-SSIM | 0.96 | 0.01 |

For the chosen quality metrics the acceptable values are:
- About 30 dB for PSNR.
- For VQM the scale is between 0 and 5, and the lower value the better.
- In the case of MS-SSIM, the values between −1 and 1 are achieved and the more similar the transmitted and received video sequence, the higher the value.

Therefore, it can be concluded that the Skype video call in the tested experimental setup was of good quality, and it can be further utilized to evaluate the proposed steganographic method. It must be also emphasized that the exact values of the quality metrics are not as important as the decrease in metric values caused by applying the information hiding technique.

**3.2.1 Skype video traffic measurements**

The measurements performed to present the basic characteristics of the Skype video call traffic were packet size and packet rate distribution.

During a typical Skype videoconference call two main streams are sent: the audio and video streams. Thus, it is first vital to identify the packets that form a video stream in the videoconference call. Fig. 3 presents the packet size distribution of a typical Skype connection.

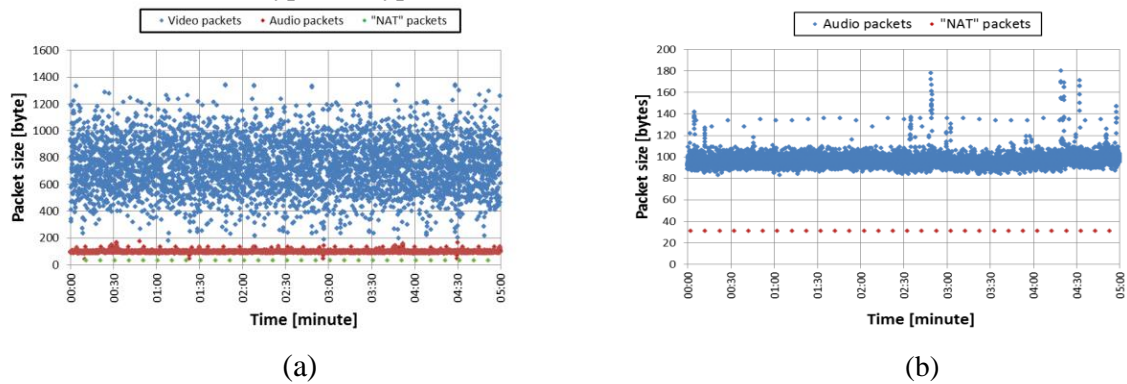

(a)            (b)

**Figure 3. Packet size distribution of a Skype (a) videoconference call, (b) voice-only call.**

Analysis of Fig. 3a reveals that during the videoconference call three different streams can be distinguished. The first stream is characterized by a packet size of 31 bytes, sent at a fixed time interval every 20 seconds. This stream is responsible for maintaining Network Address Translations (NAT) on network intermediary devices. The second stream is formed by packets with a size of about 100 bytes and the third by sizes of between 200 and 1400 bytes. If we analyse the distribution of the packet size of the voice-only call (Fig. 3b) we clearly see that the second stream is related to voice stream and the third to video traffic transmission. By inspecting closely the data in Fig. 3b we can conclude that the voice packets are smaller than 180 bytes. Therefore, for the rest of the experiments we assumed that every DATA packet of a size greater than 180 bytes formed a Skype video stream.

Moreover, it was experimentally verified that the bitrate for what was assumed to be the audio stream is about 40 kbps, and for the video stream it is about 90 kbps. Additionally, the average total rate of the DATA packets was 65 pkt/s. For packets of a size less than 180 bytes, it was about 50 pkt/s and 15 pkt/s for the audio and video stream, respectively. From previous Skype measurements it is known that the audio stream is 50 pkt/s ([7], [1]), thus confirming the correct selection of the DATA packet size for distinguishing the data streams. This also proves that each video frame is carried in a single packet in the video stream – hence the packet rate is equal to the frame rate (15 fps).

Next, we wanted to determine how Skype video traffic reacts when intentional packet losses are introduced. This allowed us to establish how Skype compensates for losses. The results of the experiments are presented in Fig. 4; for these measurements the loss level has been increased by a 1% step every 10 seconds.

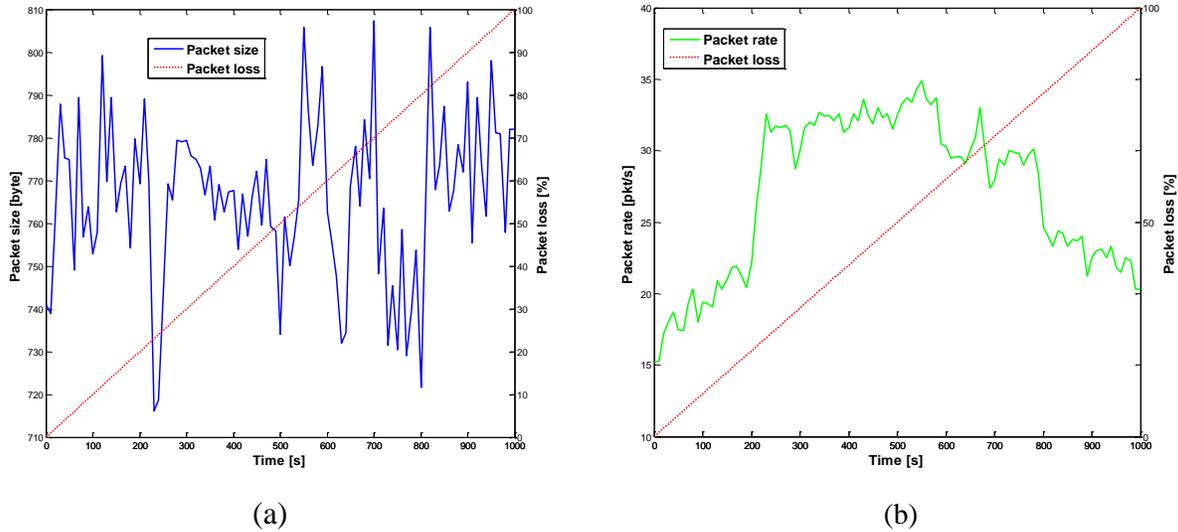

**Figure 4. Influence of packet losses on (a) the packet size, (b) the packet rate.**

From Fig. 4a, it can be concluded that the change in packet losses influences the size of the packets – the average size is 766 bytes with a standard deviation 18.67 bytes; however, it is not correlated with the packet loss level.

Also, the packets' frequency was subjected to measurements to determine whether Skype compensates for losses by increasing the data rate, as was observed for audio traffic [6] [7]. Fig. 4b confirms that the video traffic also follows this rule. When no losses are introduced, the data rate, as expected, is 15 pkt/s. With an increase in overall losses up to 20%, the frequency of the packets slowly rises to 20 pkt/s. Then, from 20% to 80% of losses the data rate increases to about 30 pkt/s, and then it drops back again to about 20 pkt/s. The observed behaviour confirms the results reported in another research on Skype video traffic [22], where it was noted that for a loss level of 16% the data rate increased by about 60%.

If the size of the packets is not correlated with the loss level and the data rate, then it is most probable that the packets are intentionally duplicated and sent to the receiving side to compensate for the elevated loss level. Our measurement revealed that part of the packets were sent with a very small time interval of a few milliseconds (while the expected interval for 15 pkt/s is about 66.6 milliseconds due to video codec frames generation). It is also worth noting that the neighbouring packets' size varies by only 5 or 6 bytes. The difference in size can be reasonably explained: the larger packet probably contains information about the replicated content.

From the designed information hiding methods perspective, the data rate fluctuations are a very important characteristic to consider. The difference in packet frequency can be utilized as an indicator for steganalysis (detection) purposes. Therefore, the next step is to analyse the number of probable packet duplicates for different loss levels.

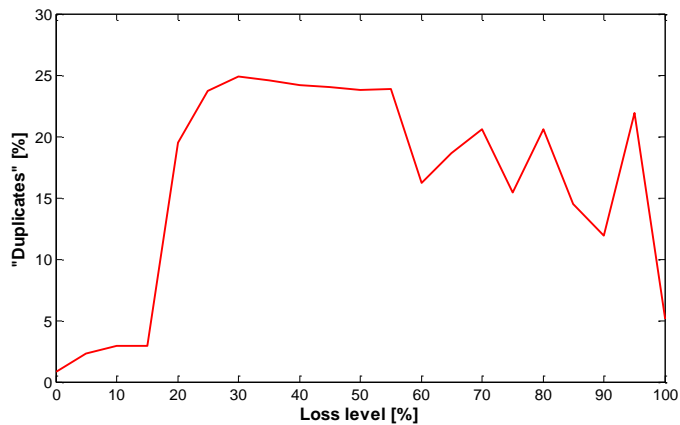

**Figure 5. Probably duplicated packets for different loss levels.**

In Fig. 5 we can observe that the number of probably replicated packets (the red curve) significantly increases when the 15% loss threshold is reached, which is in line with the trend in Figure 4b. The total number of "duplicates" is about 0.8% of all Skype video traffic when no losses are introduced.

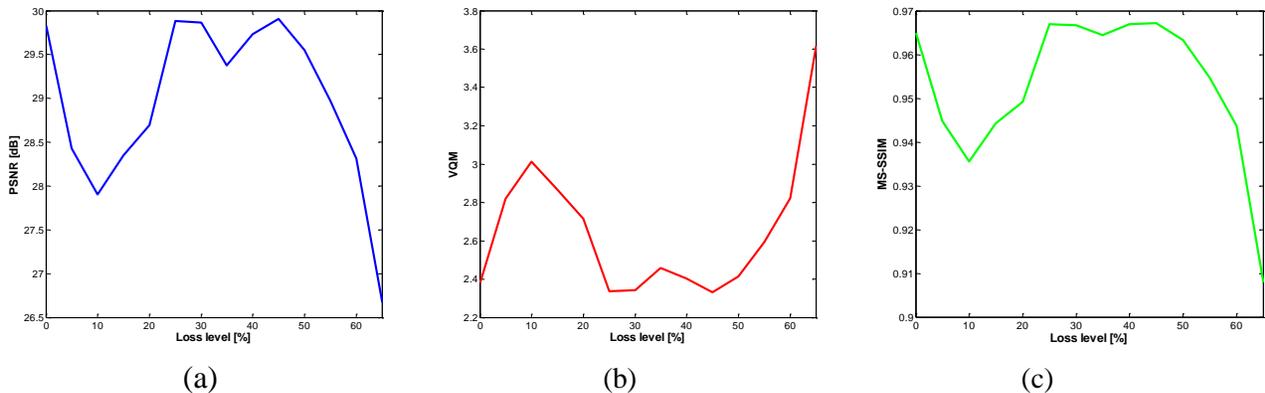

(a)　　　　　　　　　　　　　　　(b)　　　　　　　　　　　　　　　(c)

**Figure 6. Quality of the received video signal for different packet loss levels: (a) PSNR, (b) VQM, (c) MS-SSIM.**

The last part of the study is the evaluation of how the resultant quality of the video signal changes with the increase in packet losses. Fig. 6 presents the experimental results for the following metrics: PSNR, VQM and MS-SSIM for losses in range 0–65%. It should also be noted that for higher loss levels, it was not possible to synchronize the transmitted and received video signal due to its significant degradation.

PSNR (Fig. 6a) first decreases for losses in range 5–20%; then, it rises until a 50% loss level is reached, and after that it decreases significantly. The increase of PSNR at about 20% of losses is correlated with the increase in packet frequency and the utilization of probable duplicated packets (confirm by Fig. 4b). This result is in line with a previously performed measurement study [22], where the authors observed a significant degradation of video signal at 8%, which is similar to PSNR results obtained at the 10% loss level.

As expected, similar behaviour as that of PSNR was observed for other quality metrics: VQM and MS-SSIM (Fig. 6b and 6c, respectively). Therefore, it can be concluded that from the transmission quality perspective the worst packet loss levels are 10% and above 45%, and they should be avoided for the purposes of the steganographic method. It is better to keep the losses below 5% or from 20% to 45%. However, it must be also noted that for the latter the data rate increases, which is not beneficial from an undetectability point of view.

Another reason for the quite good video quality is relates to the way in which intentional losses were introduced. In this paper, until the level of 50% losses is reached only every second packet is subject to dropping. This allows the burst losses to be omitted and preserves the transmission quality. Above 50%, this rule, obviously, cannot be fulfilled. This also explains why the quality of between 20% and 50% is almost as good as introducing losses at a very low level (<1%); since there are a lot of probably duplicated video packets in this range, dropping them does not significantly affect video quality.

From the conducted measurements of Skype video traffic presented above for the proposed information hiding method, the most important requirement is to establish a baseline for the "clean" connection. This baseline can be later treated as a reference and can be compared with steganographically modified traffic. The three main and most important characteristics of the unmodified Skype video call are the following: the average data rate is 15.24 pkt/s with a standard deviation of 0.28, the average packet size is 741 bytes and about 0.8% of total number of packets are probable duplicates.

## 4.  Proposed method and its implementation

### 4.1 Threat model
YouSkyde is designed to utilize Skype encrypted video streams to enable clandestine communication. The secret data exchange can be realized in two scenarios between (S1) two Skype users (i.e. they use their own call for steganographic purposes) or (S2) a secret data transmitter and receiver that utilizes an existing third-party Skype call (in this case the original caller and callee are not aware of the information hiding procedure).

For both scenarios we can assume the presence of the global, active warden that is able to monitor Skype traffic for every videoconference call on every link in the network. However, we further assume that the warden is not a malicious one, since it would not modify the traffic blindly as this could also punish innocent Skype users.

**4.2 The method and its implementation**
The proposed method, YouSkyde, is based on the introduction of intentional losses to the Skype video traffic by utilizing the payloads of these packets to carry encrypted secret data. When the Skype traffic measurement results provided in Section 3 are considered, it turns out to be a promising approach.
We decided to pursue three different variants of this method:
- **YouSkyde variant 1 (YSv1):** 45% of losses are intentionally introduced to the video stream by utilizing their payloads to carry secret data. In this case we can expect a high steganographic bandwidth but still with good transmission quality; however, the detection could be trivial if traffic monitoring is utilized because of the resulting increase in the packets' frequency for this loss level.
- **YouSkyde variant 2 (YSv2):** only packets identified as probable duplicates (see Section 3.2.1 for details) are utilized for the purposes of the information hiding method. The resulting steganographic bandwidth is not going to be the highest one; however, it is expected that the impact on the video quality will be the lowest among the variants (as losing duplicates should not influence the quality too much). On the other hand, if the warden knows that the duplicates are utilized for steganographic purposes, then this variant is potentially easily preventable (e.g. by modifying only the duplicates).
- **YouSkyde variant 3 (YSv3):** losses are introduced at a very low level and the corresponding packets' payloads are utilized for clandestine communication purposes. The resulting steganographic bandwidth would be lower than for YSv1; however, the video quality is expected to be good and the detection or prevention of the method should be hard because: *(i)* the packets chosen for steganographic purposes are selected randomly; *(ii)* the resulting network traffic should not differ from the typical Skype videoconference call. The most suitable packet loss level for this variant will be established on the basis of the additional traffic measurement study. In the situation where the third party video call is utilized for information hiding purposes (see subsection 4.1 for details), we will also consider the average packet losses on the Internet to be sure that by applying the method we will not cross this value.

As the YouSkyde proof-of-concept prototype has been implemented, all three variants indicated above can be evaluated.

**4.2.1 Secret data transmitting side**
Fig. 7a presents state diagrams for the main algorithm responsible for the prototype's functioning and the control of the level of introduced losses. The algorithm works as follows: first only the DATA packets are selected, then the chosen variant of the YouSkyde is realized by verifying whether the size of the packet is greater than 180 bytes (for YSv1 and YSv3), or whether it is a "duplicate" (YSv2). In the next step, whether or not the maximum level of the introduced losses has been crossed is evaluated. If not, then the subprocess which governs the preparation of the secret data and their proper injection into the chosen packets' payload is started (Fig. 7b). The most important part of it is related to the generation of the START message that carries only binary ones and the STOP message that consists only of binary zeros. These two messages allow the signal to be sent to the secret data receiver at the beginning and the end of the covert communication. As both these messages are encrypted using Advanced Encryption Standard (AES) cipher (with packet size as an initialization vector and a secret password known to both secret communication sides) they cannot easily be recognized by a warden monitoring network traffic.

To ensure the secret communication's reliability, as well as each packets' integrity, two identifiers are used, each 2 bytes long. The first is a secret message identifier (*SMI*) that is concatenated with each secret message placed inside the selected Skype video packet. The second is packet identifier (*PID*), which in our prototype is created by calculating a SHA-1 hash on the secret message, and then by selecting only the two rightmost bytes. This value is later inserted into the ID field of the start of message (SoM) header in a Skype video packet (see Section 3.1 for details).

Then, the secret message to be sent together with the *SMI* is encrypted (in the same manner as indicated for the START and STOP messages) and it replaces the original data that resides in the packet's payload. The AES is utilized as the same cipher is originally used in Skype. In the final step, the UDP checksum is recalculated.

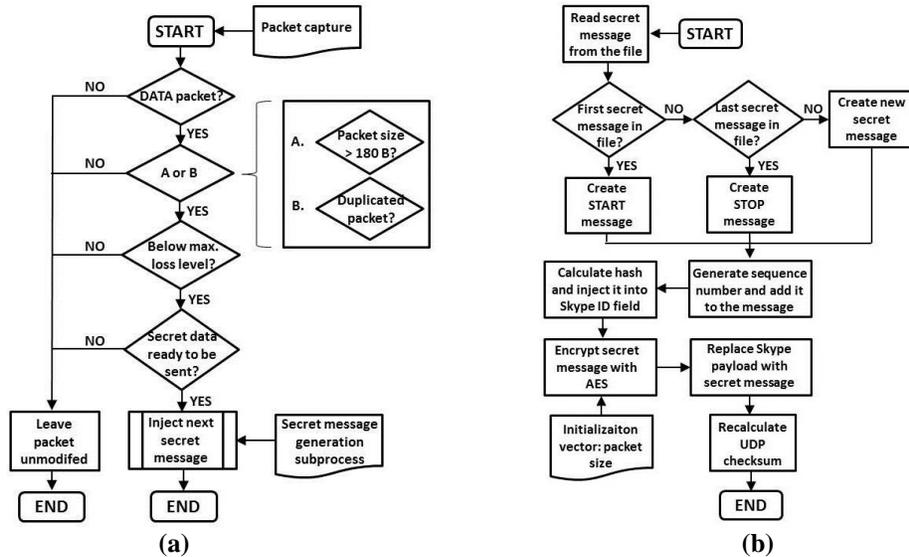

**(a)** **(b)**

**Figure 7. State diagrams of the YouSkyde transmitting side: (a) main algorithm; (b) subprocess responsible for preparing secret data before injecting them into the selected Skype video packet.**

Additionally, an additional control protocol in a hidden channel can be applied to provide increased reliability (it is sometimes called a microprotocol). One solution is to use an approach proposed by Hamdaqa and Tahvildari [8], as it can be easily incorporated into the proposed method. It provides a reliable and fault-tolerant mechanism based on a modified (k, n) threshold of a Lagrange Interpolation, and the results demonstrated by that paper prove that the complexity of the steganalysis is increased. Of course, the "cost" of the extra reliability will always be a loss of some fraction of the steganographic bandwidth.

**4.2.2 Secret data receiving side**
The secret data receiver functioning is presented as a state diagram in Fig. 8, and it is the same regardless of the YouSkyde variant chosen. First, whether the received packet is a DATA packet is verified; next, there is an attempt to decrypt the payload and the result is subjected to SHA-1. If two rightmost bytes (*PID*) are the same as those extracted from ID field of the SoM header, then it is confirmed that the packet carries secret data and that its integrity is correct. Next, the *SMI* is extracted; if it is equal to 1 and the decrypted payload looks like the START message, the covert communication is activated. Then, every packet identified as utilized for steganographic purposes is processed and analysed and the secret data is extracted. *SMI* allows packets that have been lost in the network to be identified, and the received parts of secret message to be correctly ordered. Finally, if a STOP message is received the covert communication mode is deactivated.

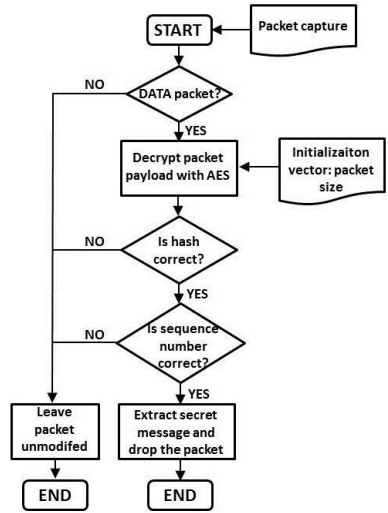

**Figure 8. State diagram of the YouSkyde receiving side.**

## 5. Experimental evaluation

As mentioned in the previous section, the presence of the probable duplicates and, as a result, the increased packet rate, puts some restraints on the designed steganographic method; that is, the level of the intentionally introduced losses must be precisely controlled so that the resulting steganographically-modified traffic is hard to distinguish from the "clean" traffic. Therefore, for the third YouSkyde variant, we must determine how the data rate changes for a low level of intentional losses.

The PingER Worldwide History Report [23] involves the measurement of over 700 sites in over 160 countries, and it monitors the different characteristics of Internet links, which includes, among others, the packet losses. In 2014, the average reported packet loss was 1.89% with a standard deviation 4.28% and with a median of 0.676. Considering this, it will be useful to measure how the packet frequency varies in the range of 0–2% of losses. The results of such an experiment are presented in Fig. 9. It is clear that the data rate up until 1% of losses does not significantly differ. However, after crossing this threshold the situation changes and the packet frequency rises considerably. Therefore, it can be concluded that from the undetectability perspective the "safe" loss level must be within the range of 0–1%. However, the best solution is to monitor the losses on a Skype video connection in real-time and adjust the level of introduced losses to the current overall packet losses for this connection so as not to cross the average packet loss for typical Internet links. This issue has been already considered for IP telephony in previous works, for example [24][25].

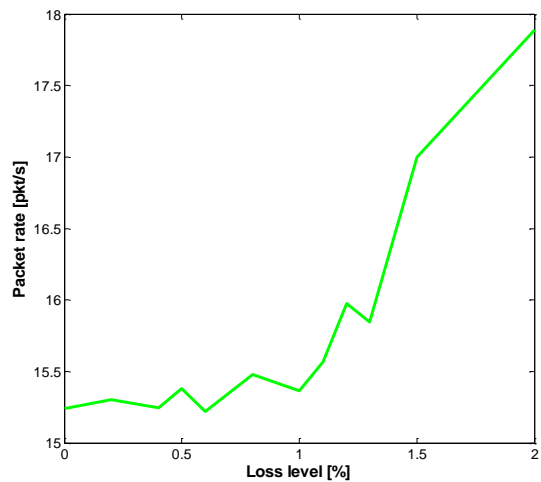

**Figure 9. Influence of packet losses in range 0–2% on data rate.**

In considering the above, for a further evaluation study of the third YouSkyde variant we chose to introduce 0.5% (YSv3_1) and 1% (YSv3_2) of intentional losses for steganographic purposes.

### 5.1 Steganographic bandwidth

The results for the steganographic bandwidth for all three variants are presented in Table 2. It must be noted that when strict undetectability is not required it is possible to covertly send with a steganographic bandwidth of 80 kbps (YSv1), which is a very high result for a steganographic method. Also, for the information hiding method that was proposed for the audio stream (SkyDe), in an analogous situation it was only possible to send about 2.8 kbps [1].

**Table 2.** Steganographic bandwidth, cost and quality metrics results for YouSkyde

| YouSkyde variant | Steganographic bandwidth [kbps] | Steganographic Cost | | | Quality metrics | | |
|---|---|---|---|---|---|---|---|
| | | PSNR | VQM | MS-SSIM | PSNR | VQM | MS-SSIM |
| 1 | 80.5 | −0.09 | −0.05 | 0.01 | 29.91 | 2.33 | 0.97 |
| 2 | 0.7 | −0.01 | −0.03 | 0.01 | 29.81 | 2.35 | 0.97 |
| 3_1 (0.5%) | 0.45 | 0.06 | −0.04 | 0.01 | 29.88 | 2.34 | 0.97 |
| 3_2 (1%) | 0.93 | 0.28 | 0.06 | 0 | 29.54 | 2.44 | 0.96 |

However, in cases where high undetectability is desired, the remaining variants should be considered, that is, the variant that exploits "duplicates" (YSv2) and the variants with low levels of introduced losses (0.5% for YSv3_1 and 1% for YSv3_2). In these cases, the resulting steganographic bandwidth is considerably lower: for YSv2 it is 0.7 kbps, for YSv3_1 it is 0.45 kbps and for YS3_2 it is 0.93 kbps. These results are, in turn, lower than the audio-based steganographic method, where it was possible to covertly send with a rate of 1.8 kbps.

The reasons for this effect will be discussed in the following subsection.

### 5.2 Undetectability

The undetectability of YouSkyde can be viewed from two perspectives: the users' and the network. When considering these two perspectives two scenarios are possible (as mentioned in Section 4.1). The method can be used between two Skype clients who are aware of the steganographic procedure so the secret data is sent in an end-to-end manner (S1), or some intermediary nodes can rely on third party Skype calls to exchange hidden information (S2). From the user's perspective, in both scenarios the secret data sender can select the highest steganographic bandwidth that does not inflict significant video signal quality degradation (YSv1), as the results in Table 1 show that each inspected variant does not degrade the transmission quality by much. This also allows another conclusion to be drawn: despite utilizing only the "duplicated" packets in YSv2, the decrease in transmission quality when compared to YSv3 is negligible.

However, when we consider the network perspective we must also take into account the traffic that is generated when YouSkyde is applied. Thus, in both scenarios S1 and S2 the introduced losses caused by the steganographic method functioning cannot be excessive because they could make overt users or the warden suspicious. Table 3 presents the experimental results for YouSkyde generated traffic and includes three traffic characteristics: the average size of the packet, the data rate and the number of probably duplicated packets. First, let us remind ourselves of the characteristics of an unmodified Skype video call: the data rate is 15.24 pkt/s with a standard deviation of 0.28, the packet size is 741 bytes and there are also about 0.8% of probable duplicates present.

**Table 3.** Experimental results for YouSkyde from network traffic perspective

| YouSkyde variant | Packet size [byte] | Data rate [pkt/s] | "Duplicates" [%] |
|---|---|---|---|
| 1 | 760.9 | 30.13 ± 0.38 | 24.04 |
| 2 | 743.9 | 15.43 ± 0.27 | 0.77 |
| 3_1 (0.5%) | 744.2 | 15.38 ± 0.31 | 0.78 |
| 3_2 (1%) | 739.4 | 15.36 ± 0.26 | 0.83 |

Analysis of the experimental results in Table 3 reveals that YSv1 can be easily detected by monitoring network traffic. Introducing 45% of losses due to YouSkyde's operation inflicts an increase in the data rate as more "duplicated" packets are generated (about 24%). Therefore, this could serve as an indicator to a warden to detect whether covert communication is taking place.

An active (but not malicious) warden could, to some extent, try to drop video packets to disturb covert communication. However, the more randomly dropped video packets there are, the worse the transmission quality; thus, when applied blindly this can significantly degrade the resulting video signal quality. Of course, in the case of YSv1 it is easier to "hit" the packets with secret data, as they constitute about 45% of all video traffic. It must be also noted that for the remaining method variants this is a considerably harder task to perform, except for YSv2. If a warden discovers that secret data is hidden in "duplicates" it can easily identify them and erase their content (as this will not impact on the quality of the transmission significantly). Therefore, from the network perspective YouSkyde will be most undetectable using YSv3 variants. Indeed, the results from Table 3 confirm this claim as the traffic characteristics are very similar, including the number of "duplicates". Yet, because YSv3_2 yields a higher steganographic bandwidth (0.93 kbps) this should be the preferred choice for YouSkyde.

The last thing to evaluate for undetectability purposes is the verification of whether the secret data that is injected into the chosen Skype video packets can be distinguished from the typical content of the unmodified traffic. For this purpose, we utilized the two-sample Kolmogorov-Smirnov (K-S) test. We decided to consider the worst case scenario for YouSkyde, that is, the variant where 45% video packets are modified (YSv1). We analysed 10 YouSkyde and 10 clean connections and extracted 1000 packets from each (which in total gives 20,000 packets). The K-S test result is p = 0.592 with a standard deviation of 0.276, which proves that the cumulative distribution functions (CDFs) of unmodified and steganographic traffic cannot be distinguished (exemplary CDFs are presented in Fig. 10).

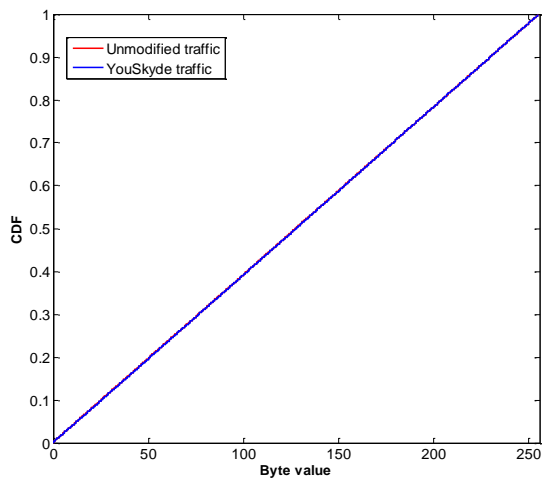

**Figure 10. Comparison of CDFs of unmodified Skype video and YouSkyde traffic for YSv1 variant (both curves are practically the same).**

## 5.3 Steganographic cost and robustness

The introduced steganographic cost depends on the variant of the YouSkyde method that is chosen. As mentioned before, the results in Table 2 reveal that, whatever the variant, its impact on transmission quality when compared with a reference Skype videoconference connection without steganography applied is negligible. However, in the case of YSv1 the steganographic cost can expressed also as an elevated level of probable duplicated packets (Table 3), as more than 24% of them are present, while for a "clean" connection it is about 0.8%.

From a robustness perspective, each variant of YouSkyde utilizes communication integrity preserving mechanisms (see Section 4.2.1 for details). Therefore, if some losses occur, for instance, due to network functioning, they will only affect the achieved steganographic bandwidth but will not break the covert communication. Missing fragments of secret data can be identified as signalled back to the secret data sender.

## 6. Conclusions and future work

In this paper, a new steganographic method for a Skype videoconference connection called YouSkyde is proposed. From the Skype network traffic perspective, it is based on introducing intentional losses into the video stream by replacing the payload of selected packets with encrypted secret messages. This approach was proved to be successful by conducting a measurement study on real Skype video traffic and performing performance evaluation on a YouSkyde prototype that has been implemented. Of the three different variants of YouSkyde considered, it turned out that the best approach is to utilize about 1% of packets from the video stream for steganographic purposes. This results in reasonably high steganographic bandwidth of 0.93 kbps and negligible transmission quality degradation within the terms of statistical undetectability.

Surprisingly, despite the higher bitrate of the Skype video stream (about 90 kbps) than the audio stream (about 40 kbps) the resulting steganographic bandwidth turned out to be about 50% lower than a previously proposed SkyDe method [1] for Skype audio streams that relied on selecting and replacing the payloads of the packets carrying silence. This is probably caused mainly by the presence of duplicated video packets that influence the resultant data rate, making introduced intentional losses easily detectable if applied excessively.

Future work will include focusing on developing an effective YouSkyde detection scheme. One research direction that can be pursued in this vein is the utilization of encrypted traffic classification schemes, as in [26] or [27], to try to distinguish encrypted video signals from encrypted secret messages.